\documentstyle[12pt,cite,epsfig]{article}

\renewcommand{\today}{23 June, 1997}

\newcommand{\nc}{\newcommand}
\nc{\be}{\begin{equation}}
\nc{\ee}{\end{equation}}
\nc{\bea}{\begin{eqnarray}}
\nc{\eea}{\end{eqnarray}}
\nc{\beas}{\begin{eqnarray*}}
\nc{\eeas}{\end{eqnarray*}}
\nc{\noi}{\noindent}
\nc{\sD}{\not \! \! D}
\nc{\s}[1]{\not \! #1}
\nc{\non}{\nonumber}
\nc{\bb}{\bibitem}
\nc{\lf}{\left}
\nc{\mb}[1]{\makebox[#1]{}}
\nc{\pa}{\partial}
\nc{\sA}{\not \! \! A}
\nc{\newsec}[1]{\section{#1}\mb{0.5cm}}
\nc{\h}{\frac{1}{2}}
\nc{\ra}{\rightarrow}
\nc{\la}{\leftarrow}
\nc{\ep}{$e^+e^-\ra\pi^+\pi^-\;$}
\nc{\epp}{$e^+e^-\ra\pi^+\pi^0\pi^-\;$}
\def\mathunderaccent#1{\let\theaccent#1\mathpalette\putaccentunder}
\def\putaccentunder#1#2{\oalign{$#1#2$\crcr\hidewidth
\vbox to.2ex{\hbox{$#1\theaccent{}$}\vss}\hidewidth}}

\nc{\ti}{\mathunderaccent\tilde}
\nc{\M}{{\cal M}}
\nc{\rw}{$\rho\!-\!\omega\;$}

\begin{document}
\thispagestyle{empty}
\begin{flushright}
ADP-97-21/T258 \\
UK/97-14 \\
hep-ph/9707404

\end{flushright}
\begin{center}
{\large{\bf Near-threshold Isospin Violation in the Pion Form Factor\\
from Chiral Perturbation Theory}} \\
\vspace{1.5 cm}
H.B.\ O'Connell$^{a,b}$, K. Maltman$^{a,c,d}$, A.W.\ Thomas$^{a,c}$
and A.G.\ Williams$^{a,c}$ \\
\vspace{.5cm}
$^{a}${\it
Department of Physics and Mathematical Physics,\\
University of Adelaide 5005, Australia } \\
\vspace{.2cm}
$^{b}${\it
Department of Physics and Astronomy, University of Kentucky,\\
Lexington, KY 40506, USA}\footnote{Present address.}\\
\vspace{.2cm}
$^{c}${\it 
Special Research Centre for the Subatomic Structure of Matter,\\
University of Adelaide 5005, Australia } \\
\vspace{.2cm}
$^{d}${\it Mathematics and Statistics, York University, 4700 Keele St., \\
North York, Ontario, Canada M3J 1P3}\footnote{Permanent address.}\\

\vspace{1.2 cm}
\today
\vspace{1.2 cm}
\begin{abstract}

We examine the isospin violation in the timelike pion form factor
near threshold at next-to-leading order in the chiral expansion
using the techniques of Chiral Perturbation Theory.  This
next-to-leading order contribution contains the first 
nonvanishing isospin violation. This isospin violation is found
to be very small near threshold.  In particular, the isospin
violation at threshold is found to be of order $10^{-4}$, which
should be compared with the few percent level seen in the vector meson
resonance region.
\end{abstract}
\end{center}
\vspace{2.5cm}
\begin{flushleft}
E-mail: {\it hoc@ruffian.pa.uky.edu;
athomas, awilliam@physics.adelaide.edu.au} \\

Keywords: vector mesons,
pion form factor, meson mixing, isospin violation.\\
PACS: 11.30.Hv, 11.30.Rd, 12.39.Fe, 12.40.Vv, 13.40.Gp

\vfill
{\it }
\end{flushleft}

\newpage

\section{Introduction}
Chiral Perturbation Theory (ChPT) seeks to produce, in a
model independent way, a completely general
low energy effective hadronic field theory, using as input
only the underlying symmetries and pattern of symmetry breaking of the
initial QCD Lagrangian.  
The principle symmetry used in this construction is
chiral symmetry (ChS). ChS is the simultaneous requirement of isospin symmetry
and helicity conservation, i.e. SU$(2)_L\otimes$SU$(2)_R$. Having $m_u=m_d\neq
0$ violates helicity conservation but not isospin symmetry. In the real world
we have $m_u\neq m_d\neq 0$ and both symmetries are broken by nonzero
current quark masses. 
Since $m_u$ and $m_d$ are small on the hadronic scale the violation of chiral
symmetry is small. Isospin is also explicitly violated by electromagnetic and
weak interactions.  The systematic nature of ChPT then provides a
model-independent method for examining isospin breaking
in the regime of applicability of the method. 

The pion form-factor, $F_\pi(q^2)$, was one of the first quantities calculated
beyond leading order in the chiral expansion
using ChPT \cite{GL}. 
However, this one-loop treatment assumed $m_u=m_d$. In this
note, we extend this treatment 
to the case with $m_u\neq m_d$, following the work of
Maltman \cite{kim}. The previous calculations of $F_\pi(q^2)$ are
briefly reviewed and the isospin-violating calculation is then discussed in
detail. It should be noted that the calculation involves
simultaneously expanding in three small parameters, 
$q^2$, $\alpha_{EM}$ and $m_u-m_d$.
We shall work to first order in each of these.

\section{An introduction to ChPT}
\mb{.5cm}
Although there are many excellent reviews of ChPT \cite{chrev}, to keep this
discussion relatively self contained, we present a short summary of the
approach. 
This will also be useful in showing how we set up the calculation.
The
basic idea is to take the known symmetries of QCD and reproduce them in a
low-energy meson theory. Thus we start with the QCD Lagrangian given 
by\footnote{To work with this, one needs to remove the unphysical 
gauge degree of
freedom which is usually accomplished by adding a gauge fixing term
to Eq.~(\ref{QCD}), however this is not important for our discussion 
and will be omitted.}
\be
{\cal L}^{\rm QCD}=\sum_f \bar{\psi}(x)(i\sD -m_f)\psi(x)-\frac{1}{4}
F_{\mu\nu}^a
F^{a\mu\nu}.
\label{QCD}
\ee
In the chiral limit the quark masses are zero and the fermion fields, 
$\psi$, can be split into left and right handed
helicity components,
\be
\psi_{L,R}=(1\pm\gamma_5)\psi.
\ee
These transform {\em independently} under the chiral transformation,
\be
\psi_{L,R}\ra e^{i\alpha\gamma_5}\psi_{L,R},
\ee
leaving Eq.~(\ref{QCD}) unchanged. Massless QCD is then said to be {\em
chirally symmetric}.
These transformations can then be generalised to separate left and right handed
transformations rather than just the single $e^{i\alpha\gamma_5}$
transforming both fields. In this case we have
\be
\psi_{L,R}\ra U^{L,R}\psi_{L,R}
\ee
where $U^L$ and $U^R$ are unitary $N^f\times N^f$ matrices, $N^f$ being the 
number of flavours. One normally only considers the up, down and strange 
quarks, for reasons that will become apparent later. The heavier
quark flavours play no dynamical role in the region of interest and do not
need to be explicitly included. If strange quarks are included the flavour
symmetry changes from SU(2)$_{\rm flavour}$ to SU(3)$_{\rm flavour}$
and the chiral symmetry group is then
SU$(3)_L\otimes$SU$(3)_R$.

Now, of course, the quarks do have mass, but since the $u$, $d$ and $s$
masses are small, SU(3)$_L\otimes$SU(3)$_R$ 
should be an {\em approximate} symmetry of QCD, and we expect it
to have some relevance to the way the theory works, and provide a guide in our
construction of a meson theory.  To construct this meson theory, we consider
the QCD generating functional, in the presence of
external left-hand vector, right-hand vector, scalar
and pseudoscalar sources $l_\mu$, $r_\mu$, $s$ and $p$,
\be \exp [iW[l_\mu,r_\mu,s,p]]=\int[{\cal
D}\psi][{\cal D}\bar{\psi}] [{\cal D}G^a_\mu]\exp\left[i\int d^4x{\cal L}_{\rm
QCD}(l_\mu,r_\mu,s,p)\right].
\ee 
The sources for the left and right
handed vector currents are decomposed into their flavour
octet components via
\be l_\mu\equiv
l_\mu^a\lambda^a/2,\;\;r_\mu\equiv r_\mu^a\lambda^a/2,
\ee 
where the $\lambda^a$ are the Gell-Mann matrices that make up the generators
of SU(3).
The sources for
the scalar and pseudoscalar ``currents'' are similarly decomposed as
\be
s=s_0+s^a\lambda^a/2,\;\;p=p_0+p^a\lambda^a/2,  
\ee 
where, for convenience, one usually includes a source term for the
singlet currents.
If we define ${\cal
L}^0_{\rm QCD}$ to be the massless QCD Lagrangian (Eq.~(\ref{QCD}) with
$m_f=0$) the full massless QCD Lagrangian, now with sources, can be written, 
\be {\cal
L}_{\rm QCD}(l_\mu,r_\mu,s,p)={\cal L}^0_{\rm QCD}-\bar{q}_L\gamma^\mu l_\mu
q_L-\bar{q}_R\gamma^\mu r_\mu q_R-\bar{q}(s-i\gamma_5p) q. 
\label{sourceqcd}
\ee 
Defining vector and axial vector current sources through
\be
l^a_\mu=v_\mu^a-a^a_\mu,\,\,\,r^a_\mu=v_\mu^a+a^a_\mu,
\ee
we can rewrite Eq.~(\ref{sourceqcd})
in terms of these vector and axial vector sources 
\be {\cal
L}_{\rm QCD}(v_\mu,a_\mu,s,p)={\cal L}^0_{\rm QCD}
-\bar{q}(\s{v}+\s{a}\gamma_5)q-\bar{q}(s-i\gamma_5p)q.  
\ee   
The role of the sources is an
important one in the construction of the effective low-energy
hadronic theory since it turns out that the QCD Lagrangian
in the presence of these sources has a local symmetry (to be discussed
below) which must, therefore, also be realised in any low-energy
effective version of the theory, if that effective theory is
to correctly represent the effects of QCD.

The local symmetry mentioned above consists of the following 
simultaneous {\it local} transformations of the left- and right-handed quark 
fields and external sources:  if the left- and right-handed quark
fields are transformed via the matrices $L(x)$ and $R(x)$, respectively, 
then the external left-handed and right-handed vector sources, 
$l_\mu$ and $r_\mu$, transform as corresponding gauge fields, and the 
scalar and pseudoscalar sources as
\bea
(s+ip)(x)&\ra& R(x)(s+ip)L^{\dag}(x) 
\label{trans1} \\
(s-ip)(x)&\ra& L(x)(s-ip)R^{\dag}(x)\ .
\label{trans2}  
\eea
Under the above set of transformations, 
${\cal L}_{\rm QCD}$ has a larger {\em local} 
${\rm SU}(3)_L\otimes {\rm SU}(3)_R$ symmetry, which must be
also realised in any low-energy effective hadronic field theory
purporting to be a representation of QCD.  

For the Goldstone boson
fields, it has long been known\cite{cwz} that it is possible to
choose the pseudoscalar fields, $\pi^a\ (a=1,\cdots , 8)$,
in such a way that, with $\lambda^a$ the usual Gell-Mann matrices
and $\pi\equiv\pi^a\lambda^a$,
\bea
\pi\equiv\pi^a\lambda^a 
&=&\left(\begin{array}{ccc}
\pi^3+\pi^8/\sqrt{3} & \sqrt{2}\pi^+ &  \sqrt{2}K^+ \\
\sqrt{2}\pi^- & -\pi^3+\pi^8/\sqrt{3} & \sqrt{2}K^0 \\
\sqrt{2}K^- & \sqrt{2}\bar{K}^0 & -2\pi_8/\sqrt{3} \end{array}\right),
\label{piU}
\eea
the matrix variable, $U=\exp ({i\pi/F})$, transforms linearly under
the chiral group,
\be
U(x)\ra L(x)U(x)R^{\dag} (x)\ .
\label{fieldtransf}
\ee
The low-energy effective theory for the Goldstone boson degrees
of freedom is then to be constructed in such a way that each
term appearing in the effective Lagrangian is invariant under the
simultaneous transformation of the external sources described
above and the transformation of the pseudoscalar fields
implicit in Eq.~(\ref{fieldtransf}).  When the external scalar
source, $s$, in the effective theory, is set equal to the
quark mass matrix,
\be
s=m\equiv\left(\begin{array}{ccc}
	m_u & 0 & 0 \\
	0 & m_d & 0 \\
	0 & 0 & m_s
	\end{array}\right),
\label{mass}
\ee
then we correctly incorporate the explicit breaking of the chiral
symmetries in QCD into the low-energy effective theory with the
same symmetry breaking pattern with which this breaking occurs
in QCD.  To say this in another way, the external field method
is essentially a spurion method for incorporating the breaking
of chiral symmetry into the low-energy effective theory.  The
external left- and right-handed vector, and pseudoscalar sources
are then useful for generating the correct representations
of the corresponding hadronic currents in terms of the Goldstone
boson degrees of freedom.

Importantly for our work, the mass matrix in Eq.~(\ref{mass}) breaks more than
just chiral symmetry, it also breaks isospin symmetry (in the strong
interaction itself) if $m_u\neq m_d$.

We may now proceed to review the construction of the low-energy effective
theory for the Goldstone bosons based on the proceedure described above.
The Lagrangian is written as a series in powers of $q^2$ and/or the
quark masses, where, owing to the fact that the pseudoscalar squared-masses
are linear in the quark masses at leading order, $m_q$ counts as
${\cal O}(q^2)$.  The counting based on
this identification defines the so-called ``chiral order''.  Labelling
the terms in the effective Lagrangian by their chiral order one then
has
\be
{\cal L}=\sum_{n=1} {\cal L}_{2n},
\label{orderlag}
\ee
where the subscript denotes the chiral order.  The construction of
${\cal L}_{2n}$ is, in principle, straightforward.
At each order one simply writes down all possible terms 
invariant under the simultaneous transformations of the
matrix variable $U$ and the external sources, each such term
multiplied by a coefficient which is, of course,
not fixed by the symmetry arguments alone.  These coefficients,
called ``low-energy constants'' (or LEC's), are to be fixed by comparison 
with experiment, or estimated in some model-dependent approach
(see, for example, Ref.~\cite{li}).  
Although the resulting full effective theory necessarily has
an infinite number of terms, and hence is non-renormalisable,
to a given order in the chiral series, only a finite number
of these terms contribute\cite{weinberg79}, so that the theory
becomes effectively renormalisable.

The most general form of ${\cal L}$, at lowest order in the chiral
expansion, is then easily seen to be
\be
{\cal L}_2=\frac{F^2}{4}\langle D_\mu U^{\dag}D^\mu U+U^{\dag}\chi+
\chi^{\dag}U\rangle,
\label{lag}
\ee
where $F$ is one of the LEC's mentioned above, which has the
dimensions of mass and turns out to be equal to the
pion decay constant in the chiral limit, 
$\langle A\rangle$ denotes the trace of matrix $A$,
the covariant derivative $D_\mu U$ is defined by
\be
D_\mu U=\pa_\mu U +i[v_\mu,U]-i\{a_\mu,U\},
\label{cov}
\ee
and the source, $\chi$ by,
\be
\chi=2B_0(s-ip),
\label{chi}
\ee
where $B_0$ is another LEC, whose physical meaning turns out to
be that the quark condensate in the chiral limit is $-B_0 F^2$.

The lowest order part of the Lagrangian, Eq.~(\ref{lag}), 
produces the kinetic and mass terms for (say) 
the pion field, when we set $s$ in
Eq.~(\ref{chi}) to the quark mass matrix of Eq.~(\ref{mass}). We simply expand
the exponential of $U$ in terms of the pion field to give
\be
\frac{F^2}{4}2B_0\langle m(U^{\dag}+U)\rangle=B_0\left(-\langle m\pi^2
\rangle+\frac{1}{6F^2}\langle m\pi^4\rangle +\cdots\right).
\ee
Making the appropriate identifications gives us the well-known
Gell-Mann--Oakes--Renner relation \cite{GOR}
between the quark and meson masses
(modified to include the 
leading isospin-breaking contributions\cite{chrev})
\be
\begin{array}{cc}
m^2_{\pi^{\pm}}=(m_u+m_d)B_0, & m^2_{\pi^{0}}=(m_u+m_d)B_0-\delta+
O(\delta^2) \\
m^2_{K^{\pm}}=(m_u+m_s)B_0, & m^2_{K^{0}}=(m_d+m_s)B_0 \end{array}
\label{meson-masses}
\ee
where the second-order CSV parameter, $\delta$, is given by
\be
\delta=\frac{B_0}{4}\frac{(m_u-m_d)^2}{(m_s-m_u-m_d)}.
\ee
One can easily see from this, that if the quark masses vanish, then so do
the pseudoscalar meson masses.

Our calculation will incorporate terms at leading
and next-to-leading order in
the chiral expansion.  Following Weinberg's counting 
argument\cite{weinberg79}, this means that we must include
1-loop graphs with vertices from ${\cal L}_2$ in addition to
tree graphs with no more than one vertex from ${\cal L}_4$.
The terms involving the vertices from ${\cal L}_4$ serve to
renormalise (at this order) the divergences generated by
the loop graphs involving vertices from ${\cal L}_2$.

The explicit form of ${\cal L}_4$ was worked out by Gasser
and Leutwyler\cite{GL}, and is given by
\bea
{\cal L}_4&=& L_1\langle D_\mu UD^\mu U^{\dag}\rangle^2+ L_2\langle D_\mu
UD_\nu U^{\dag}\rangle\langle D^\mu UD^\nu U^{\dag}\rangle +L_3\langle
D_\mu UD^\mu U^{\dag}D_\nu U D^\nu U^{\dag}\rangle\non \\
&& +L_4\langle D_\mu
UD^\mu U^{\dag}\rangle\langle\chi
U^{\dag}+U\chi^{\dag}\rangle+ 
 L_5\langle D_\mu UD^\mu U^{\dag}(\chi
U^{\dag}+U\chi^{\dag})\rangle\non \\
&& +L_6\langle\chi
U^{\dag}+U\chi^{\dag}\rangle^2+L_7\langle\chi
U^{\dag}-U\chi^{\dag}\rangle^2+L_8\langle\chi U^{\dag}\chi
U^{\dag}+U\chi^{\dag}U\chi^{\dag}\rangle\non \\
&&+iL_9\langle L_{\mu\nu}
UR^{\mu\nu}U^{\dag}\rangle
+H_1\langle
R_{\mu\nu}R^{\mu\nu}+L_{\mu\nu}L^{\mu\nu}\rangle+H_2\langle\chi^{\dag}\chi
\rangle.
\label{L4}
\eea
This constitutes the complete set of linearly-independent
terms in SU(3)$\otimes$ SU(3) allowed by the relevant symmetries and of
order $q^4$ in the chiral expansion (remember that $\chi$, as defined
above, is to be considered as ${\cal O}(q^2)$ in the chiral counting).

\section{The standard ChPT treatment of $F_\pi(q^2)$}
\mb{.5cm}
The pion form factor, $F_\pi(q^2)$, is defined as the strong interaction
correction to the naive, electromagnetic prediction of the amplitude
for \ep \cite{review}.
So to obtain $F_\pi(q^2)$ we need some way to incorporate the photon into 
ChPT.  This is straightforward since the low-energy representation
of the electromagnetic current may be obtained simply by the
variation of ${\cal L}$ with respect to the corresponding external
source, $v_\mu^{EM}=v_\mu^3 + v_\mu^8/\sqrt{3}$.  Equivalently,
one may drop all external sources except for $s$ (to be set
equal to the quark mass matrix $m$) and $v_\mu^{EM}$, and
replace $v_\mu^{EM}$ by the matrix $B_\mu$ obtained by
multiplying the photon field variable $A_\mu$ by the quark
charge matrix, $Q$,
\bea
B_\mu&=&B^3_\mu+B^8_\mu,\non\\
&=&A_\mu Q\ =\ A_\mu(x)(Q^3+Q^8),
\eea
where
\be
Q=\left(\begin{array}{ccc}
2/3 & 0 & 0\\
0 & -1/3 & 0\\
0 & 0 & -1/3\end{array}\right)=\frac{e}{2}\left(\lambda_3+\frac{1}{\sqrt{3}}
\lambda_8\right)=Q^3+Q^8,
\ee
in which
case the covariant derivative reduces to 
\be
D_\mu U=\pa_\mu U +ie[A_\mu Q,U].
\label{photon-cov}
\ee
The isospin conserving ($m_u=m_d$) treatment was first performed
by Gasser and Leutwyler\cite{GL}.  The result is
\be
F_\pi^{u=d}(q^2)=1+q^2\left[\frac{2 L}{F^2}+\frac{1}{192 F^2 \pi^2} A
\right],
\label{usualff}
\ee
where $F$ is as described above (and can be related to
the pion decay constant $f_\pi =92.4$ MeV through the 1-loop expression
of Gasser and Leutwyler\cite{GL}) and
\beas
A&=&2 \ln(\mu^2 /m_\pi^2)+\ln(\mu^2 /m_K^2) -B,\\
B&=&1+2(1-4 m_\pi^2/q^2) H(q^2/m_\pi^2)+(1-4 m_k^2/q^2) H(q^2/m_K^2),
\eeas
with $\mu$ the renormalisation scale.
(The details are analogous to those for the case $m_u\not= m_d$,
which will be outlined below.)  In Eq.~(\ref{usualff}),
$L$ is the renormalised low energy constant, $L_9^r(\mu )$, 
(see Eq.~(\ref{L4})).  The full result is, of course, independent
of $\mu$, though, in evaluating our result numerically, we will work at
$\mu =m_\rho$, for which
\be L \equiv L^r_9(m_\rho) = (6.9 \pm 0.7 )\times 10^{-3}.\ee
The quantity $H$ is defined by
\beas
H(q^2/m_\pi^2)&=&-2+2\sqrt{\frac{4m_\pi^2}{q^2}-1}
 \,{\rm arccot}\;\sqrt{\frac{4m_\pi^2}{q^2}-1}
,\,\,\, 0<q^2<4m_\pi^2 \\
&=&-2+\sqrt{1-\frac{4m_\pi^2}{q^2}}
\left(\ln\left|\frac{\sqrt{1-\frac{4m_\pi^2}{q^2}}+1}
{\sqrt{1-\frac{4m_\pi^2}{q^2}}-1}\right|+i\pi\right),\,\,\,q^2>4m_\pi^2.
\eeas

\section{The ($m_u-m_d$) contribution}
\mb{.5cm}
We are now in a position to examine contributions to the $F_\pi(q^2)$ 
resulting from the quark mass difference. 
Since it is known that isospin breaking in the isovector component
of the pion form factor is ${\cal O}\left[ (m_d-m_u)^2\right]$\cite{GL},
the ${\cal O}(m_d-m_u)$ contributions are all generated by the
isoscalar ($a=8$) component of the electromagnetic current.
We obtain the low-energy representation of the isosinglet 
electromagnetic current, $J^8_\mu$, as usual, by identifying those
terms in the effective Lagrangian linear in the external source $v^8_\mu$.

\begin{figure}[htb]
  \centering{\
    \epsfig{angle=0,figure=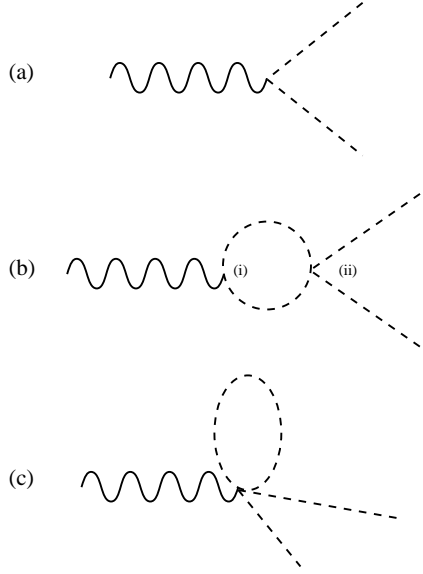,height=7.5cm}  
    }
\parbox{130mm}{\caption{The chiral contributions to 
$\gamma\ra \pi^+\pi^-$.}
\label{graphs}}
\end{figure}

Before writing down the contributions to $J_\mu^8$, it is useful
to have in mind a
picture of the graphs that will be relevant to us. These
are shown in Fig.~\ref{graphs}. We can see from these figures which
pieces of $J_\mu^8$ 
will be required.  The first such contribution is the tree-level piece
from ${\cal L}_2$, involving $\pi^+\pi^-$ 
(corresponding to Fig~\ref{graphs}.a). Second, for $M$
any meson in Eq.~(\ref{piU}) 
we need terms of the form $\bar{M}M$ in $J_\mu^8$
to give the contribution at the current vertex in Fig~\ref{graphs}.b, 
and those terms of the form $\bar{M}M\pi^+\pi^-$ 
arising from the
kinetic and mass pieces of ${\cal L}_2$ to generate the corresponding
strong vertex.  Fig~\ref{graphs}.c is generated by a
term in $J_\mu^8$ itself of the form $\bar{M}M\pi^+\pi^-$. 
The final possible contribution at this order, involving one vertex from
${\cal L}_4$, must necessarily correspond to a tree graph and hence is
of the form discussed above for Fig~\ref{graphs}(a).

It is now a straightforward algebraic exercise to obtain the
relevant contributions to the low-energy representation of
$J^8_\mu$.  {From}  ${\cal L}_2$ we find that the terms relevant to
our calculation are (the full expression is given in the Appendix)
\bea
\non
\left[ J^8_\mu\right]^{(2)}
&=&\!i\frac{\sqrt{3}}{2}\Biggl(\pa_\mu K^0\bar{K}^0-\pa_\mu\bar{K}^0K^0+
\pa_\mu K^+K^--\pa_\mu K^-K^+\Biggr)+ 
 \frac{i\sqrt{3}}{4F^2}\Biggl(\pa_\mu\pi^-\pi^+K^+K^- \non \\
&&-\pa_\mu\pi^+\pi^-K^+K^- 
+\pa_\mu K^-K^+\pi^+\pi^--\pa_\mu K^+K^-\pi^+\pi^- +
\pa_\mu\pi^+\pi^-K^0\bar{K}^0 \non \\
&&-\pa_\mu\pi^-\pi^+K^0\bar{K}^0+\pa_\mu\bar{K}^0K^0\pi^+\pi^--\pa_\mu 
K^0\bar{K}^0\pi^+\pi^-\Biggr)+{\cal O}\left( (\pi^a)^6\right).
\label{chiral-current}
\eea
We notice in
Eq.~(\ref{chiral-current}) that there is no tree-level contribution 
(Fig.~\ref{graphs}(a))
coming from ${\cal L}_2$. To calculate the vertices in
Fig.~\ref{graphs}(b) we require
the $\pi^4$ parts of the kinetic and mass terms 
of ${\cal L}_2$. These are given by 
(we assume here summation over the Lorentz indices of the 
partial derivative)
\bea
{\cal L}_{2(\pi^4)}^{\rm{KE}}\!\!&=&\!\!\frac{1}{6F^2}(2\pa\pi^+\pi^-\pa K^+
K^--\pa\pi^-
\pa K^+K^--\pi^+\pi^-\pa K^+\pa K^--\pa\pi^+\pa\pi^- K^+K^- \non \\
&&-\pa\pi^+\pi^- K^+\pa K^-+2\pi^+\pa\pi^- K^+\pa K^-+2\pa\pi^+\pi^-
K^0\pa\bar{K}^0-\pa\pi^+\pi^-\pa K^0\bar{K}^0 \non \\
&&-\pi^+\pi^-\pa K^0\pa\bar{K}^0
-\pa\pi^+\pa\pi^-K^0\bar{K}^0+2\pi^+\pa\pi^-\pa K^0\bar{K}^0
-\pi^+\pa\pi^-K^0\pa\bar{K}^0) \non \\
\non
{\cal L}_{2(\pi^4)}^{\rm mass}\!&=&\!\frac{B_0}{6F^2}
[(2m_u+m_d+m_s)\pi^+\pi^-K^+K^-
+(m_u+2m_d+m_s)\pi^+\pi^-K^0\bar{K}^0] ,
\eea
where we have written down explicitly only those contributions 
relevant to the calculation at hand.

This takes care of the contributions from ${\cal L}_2$. We must now go to
${\cal L}_4$, given in Eq.~(\ref{L4}). As it turns out, there are no
${\cal L}_4$ contributions to the low-energy representation of $J^8_\mu$,
though we might have expected a 
contribution from ${\cal L}_4$ to Fig.~\ref{graphs}(a). Usually in ChPT
such a term is responsible for removing the divergences 
(as well as the unphysical dependence on the scale, $\mu$) associated with
the loops of  Fig.~\ref{graphs}(b) and (c). Thus,
the loop graphs themselves must combine to give a finite answer.

We are now in a position to construct the Feynman amplitudes associated with
the graphs of Fig.~\ref{graphs}.  The problem
is completely standard (a good discussion of the relevant loop
integrals can be found in, for example Ref.~\cite{hollik}).
We obtain the amplitude for $A^8\ra\pi^+\pi^-$, $\M_\mu$ defining
the associated form-factor by
\be
\M_\mu=-ie(p_+-p_-)_\mu F_\pi^8(q^2).
\ee
The calculation of the amplitude is described in detail in the 
appendix, so we merely present the result for the form-factor here
\be
F_\pi^8(q^2)=-\frac{\sqrt{3}}{4F^2}\left[\frac{1}{96\pi^2}q^2\ln
\frac{m^2_{K^\pm}}{m^2_{K^0}}-\frac{1}{960\pi^2}q^4\left(\frac{1}{m^2_{K^\pm}}
-\frac{1}{m^2_{K^0}}\right)\right].
\label{firstone}
\ee
Using Eq.~(\ref{meson-masses}) we can rewrite 
this form-factor in terms of the quark masses,
\be
F_\pi^8(q^2)=-\frac{\sqrt{3}}{4F^2}\left[\frac{1}{96\pi^2}q^2\ln
\frac{m_u+m_s}{m_d+m_s}-\frac{1}{960\pi^2}q^4\left(
\frac{m_d-m_u}{B_0(m_u+m_s)(m_d+m_s)}\right)\right].
\label{secondone}
\ee
It is then easily seen that the contribution to the pion form-factor
from $J_\mu^8$ vanishes when $m_u=m_d$ as required in the
isospin limit.

\section{Discussion}
\mb{.5cm}
Setting $q^2=4m_\pi^2$ in Eq.~(\ref{firstone}) reveals a surprisingly small
(${\cal O}(10^{-4})$) isospin violation in the next-to-leading order
correction to the leading order
expression for $F_\pi(q^2)$ as given in
Eq.(\ref{usualff}).  
Hence we see that in the pion electromagnetic
form factor near threshold, the first nonvanishing
isospin violation encountered in the chiral expansion is much smaller
than the few percent level seen in the vector meson
resonance region.
The first response to this might be to assume that
$J_\mu^8$ contributes little to the pion form-factor in the low $q^2$ region
relevant to ChPT. While this may well be true, one should
bear in mind that certain features of our results imply that 
the higher order contributions to the isoscalar form factor might
not be negligible.
Basically, the low energy constants of Eq.~(\ref{L4}) are
the result of ``integrating out" the heavy resonances in an extended Lagrangian
that includes the vector mesons as well as the pseudoscalar octet.  Thus, in
any calculation where the low energy constants are absent, such as this one,
the effects of the vector resonances 
have not yet been included to the order considered
in the chiral expansion. As the isospin violation
in $F_\pi(q^2)$, at least in the resonance region, is
known to be due largely to the $\omega$ we would expect the corresponding
$\omega$ dominated LEC's to 
to play an important  role in isospin breaking
even near threshold. We can compare the situation with that of the decay
$\eta\ra \pi^0\gamma\gamma$ where the one loop ChPT prediction \cite{etadecay}
is approximately 170 times smaller than the experimental result.  The ${\cal
O}(q^6)$ contributions then bring the ChPT result into satisfactory accord with
experiment. Maltman finds a similar situation in his calculation for the mixed
current correlator $\langle0|T( V_\mu^3 V_\nu^8)|0\rangle$.  
Isospin violation is most visible in the pion form-factor data around the
$\omega$ pole where we determine that the $\omega$ contributes with a strength
$\sim 3\%$ that of the $\rho$. Although one cannot probe the resonance
pole region using ChPT, it
would thus be very interesting to see a similar two loop study of the pion
form-factor including isospin violating effects.  

\vspace{2.5cm}
\begin{center}
{\large \bf{Acknowledgements}}
\end{center}
This work was supported by the Australian Research Council
and the U.S. Department of Energy under Grant DE-FG02-96ER40989.

\appendix

\section{Chiral Perturbation Theory expressions}
\mb{.5cm}
The full expression for the current, $J_\mu$ is given by
\bea
J^8_\mu&=&( i/2 \sqrt{3} {\pa}{K^0} {\bar{K^0}} +  i/2 \sqrt{3} {\pa}{K^+} 
{K^-} -  i/2 \sqrt{3} {\pa}{K^-} {K^+} 
\non \\ &&- ( i {\pa}{K^0} {\bar{K^0}} {K^-}
{K^+})/(\sqrt{3} F^2) 
 - ( i {\pa} {K^+} {K^-}^2 {K^+})/(\sqrt{3}
F^2) \non\\
&&+ ( i {\pa}{K^-} {K^-} {K^+}^2)/(\sqrt{3} F^2) -  i/2 \sqrt{3}
{\pa}{\bar{K^0}} {K^0} \non \\ 
&&- ( i {\pa}{K^0} {\bar{K^0}}^2
{K^0})/(\sqrt{3} F^2) - ( i {\pa}{K^+ } {\bar{K^0}} {K^-} {K^0})/(\sqrt{3} F^2)
 \non \\ &&  +  ( i {\pa}{K^-} {\bar{K^0}} {K^+} {K^0})/(\sqrt{3} F^2) + ( i
{\pa}{\bar{K^0}} {K^-} {K^+} {K^0})/(\sqrt{3} F^2) \non \\
&&+ ( i {\pa}{\bar{K^0}}
{\bar{K^0}} {K^0}^2)/(\sqrt{3} F^2) \non \\ 
&&+ ( i/2 \sqrt{ 3/2} {\pa}\pi_3
{\bar{K^0}} {K^+} {\pi^-})/F^2 -     ( i/4 \sqrt{3} {\pa}{\pi^+} {K^-} {K^+}
{\pi^-})/F^2 \non \\ && + ( i/4 \sqrt{3} {\pa}{\pi^+} {\bar{K^0}} {K^0}
{\pi^-})/F^2 + ( i/4 \sqrt{3} {\pa}{\pi^-} {K^-} {K^+} {\pi^+})/F^2 \non \\ 
&&-
( i/4 \sqrt{3} { \pa}{\pi^-} {\bar{K^0}} {K^0} {\pi^+})/F^2 - ( i/2 \sqrt{3/2}
{\pa}\pi_3 {K^-} {K^0} {\pi^+})/F^2 \non \\ &&- ( i/4 {\pa}{K^0} {\bar{K^0}}
{\pi^-} {\pi^+})/(\sqrt{3} F^2) - ( i/4 {\pa}{K^+} {K^-} {\pi^-}
{\pi^+})/(\sqrt{3} F^2) \non \\ 
&&+ ( i/4 {\pa}{K^-} {K^+} {\pi^-}
{\pi^+})/(\sqrt{3} F^2) + ( i/4 {\pa}{\bar{K^0}} {K^0} {\pi^-}
{\pi^+})/(\sqrt{3} F^2) \non \\ 
&&- ( i/2 \sqrt{3/2} {\pa}{\pi^-} {\bar{K^0}}
{K^+} \pi_3)/F^2 +    ( i/2 \sqrt{3/2} {\pa}{\pi^+} {K^-} {K^0} \pi_3)/F^2 \non
\\ &&- ( i/8 {\pa}{K^0} {\bar{K^0}} \pi_3^2)/(\sqrt{3} F^2) - ( i/8 {\pa}{K^+}
{K^-} \pi_3^2)/(\sqrt{3} F^2)\non \\
&& + ( i/8 {\pa}{K^-} {K^+} \pi_3^2)/(\sqrt{3} F^2)
 \non \\ 
&&  +  ( i/8 {\pa}{\bar{K^0}} {K^0} \pi_3^2)/(\sqrt{3} F^2) - ( i/2
{\pa}{K^+} {\bar{K^0}} {\pi^-} \pi_8)/(\sqrt{2} F^2)  \non \\ 
&&  +  ( i/2
{\pa}{\bar{K^0}} {K^+} {\pi^-} \pi_8)/(\sqrt{2} F^2) - ( i/2 {\pa}{K^0} {K^-}
{\pi^+} \pi_8)/(\sqrt{2} F^2) \non \\ 
&&+ ( i/2 {\pa}{K^-} {K^0} {\pi^+}
\pi_8)/(\sqrt{2} F^2) + ( i/4 {\pa} {K^0} {\bar{K^0}} \pi_3 \pi_8)/F^2 - \non
\\ &&   ( i/4 {\pa}{K^+} {K^-} \pi_3 \pi_8)/F^2 + ( i/4 {\pa}{K^-} {K^+} \pi_3
\pi_8)/F^2 - ( i/4 {\pa}{\bar{K^0}} {K^0} \pi_3 \pi_8)/F^2 \non \\ 
&&- ( i/8
\sqrt{3} {\pa}{K^0} {\bar{K^0}} \pi_8^2)/F^2  \non \\ 
&&  -  ( i/8 \sqrt{3}
{\pa}{K^+} {K^-} \pi_8^2)/F^2 + ( i/8 \sqrt{3} {\pa}{K^-} {K^+} \pi_8^2)/F^2
\non \\
&& +( i/8 \sqrt{3} {\pa}{\bar{K^0}} {K^0} \pi_8^2)/F^2) +\cdots
\label{full-current} \eea
where we have written down explicitly only those terms required in
our calculation.

\section{Useful Integrals}
\mb{.5cm}
The following integrals are treated in some detail in Refs.~\cite{hollik,GK}.
However, Golowich and Kambor expand the expressions in powers of $q^2$ as
required for ChPT.

Let us define the one-point integral, in $D=4-\epsilon$ dimensions
\be
\frac{i}{16\pi^2}\mu^{D-4}
A(m^2)\equiv\int\frac{d^Dk}{(2\pi)^D}\frac{1}{k^2-m^2},
\ee
where $\mu$ is an arbitrary mass scale required to to keep the 
action ($\int d^Dx {\cal L}_{\rm int}$) dimensionless.
Evaluating $A(m^2)$ gives us
\be
A=m^2\left(\Delta-\ln \frac{m^2}{\mu^2}+1\right)+O(\epsilon)
\label{aeq}
\ee
where
\be
\Delta=\frac{2}{\epsilon}-\gamma+\ln 4\pi.
\label{delta}
\ee
For convenience we define the quantity
\be
\sigma=\frac{i}{16\pi^2}.
\label{sigma}
\ee
The higher-point functions are, of course, more complicated, but are related
in such a way that one can simplify expressions before calculating them 
explicitly.
\bea
\sigma\mu^{D-4}B(q^2,m^2)&\equiv&\int\frac{d^Dk}{(2\pi)^D}\frac{1}{
(k^2-m^2)((k+q)^2-m^2)}\\
\sigma\mu^{D-4}B_\mu(q^2,m^2)
&\equiv&\int\frac{d^Dk}{(2\pi)^D}\frac{k_\mu}{(k^2-m^2)((k+q)^2-m^2)}\\
\sigma\mu^{D-4}B_{\mu\nu}(q^2,m^2)
&\equiv&\int\frac{d^Dk}{(2\pi)^D}\frac{k_\mu k_\nu}{(k^2-m^2)((k+q)^2-m^2)}.
\eea
{From} simple Lorentz covariance, we can rewrite these as,
\bea
B_\mu(q^2,m^2)&=&q_\mu B_1(q^2,m^2) \\
B_{\mu\nu}(q^2,m^2)&=&q_\mu q_\nu B_{21}(q^2,m^2)
+g_{\mu\nu}B_{22}(q^2,m^2)
\eea
The functions $B_{21}$ and $B_{22}$ can be written in terms of $A(m^2)$ 
and $B(q^2,m^2)$
\cite{hollik}
\bea
B_{21}(q^2,m^2)&=&\frac{1}{3q^2}\left[A+(q^2-m^2) B-m^2+\frac{q^2}{6}\right]
\label{b21}\\
B_{22}(q^2,m^2)&=&\frac{1}{6}\left[A+(2m^2-\frac{q^2}{2})B+2m^2-\frac{q^2}{3}
\right].\label{b22}
\eea
$A(m^2)$ is given in Eq.~(\ref{aeq}) and $B(q^2,m^2)$ is given by
\be
B(q^2,m^2)=\Delta-\int_0^1dx\ln\frac{x(x-1)q^2+m^2}{\mu^2}.
\ee
We see from Eq.~(\ref{delta}) that $B(q^2,m^2)$ is divergent. Not only
that but,
as Golowich and Kambor \cite{GK} point out, it should be expanded in
powers of
$q^2$ or otherwise our use of it in ChPT will not be consistent. To do this
they define
\bea\non
\overline{B}(q^2,m^2)&\equiv&B(q^2,m^2)-B(0,m^2)\\ \non
&=&-\int_0^1dx\ln\left(1-x(1-x)\frac{q^2}{m^2}\right)\\
&=&\frac{1}{6}\frac{q^2}{m^2}+\frac{1}{60}\frac{q^4}{m^4}+...
\eea
We note now that 
\be
B(0,m^2)=\frac{\pa}{\pa m^2}A(m^2) 
=\frac{A(m^2)}{m^2}-1
\ee
We therefore rewrite Eqs.~(\ref{b21}) and (\ref{b22}) 
using
\be
B(q^2,m^2)=\overline{B}(q^2,m^2)+\frac{A(m^2)}{m^2}-1.
\ee
We arrive at \cite{GK}
\bea
B_{21}(q^2,m^2)&=&\frac{1}{3}\left[\left(1-\frac{m^2}{q^2}\right)
\overline{B}+\frac{A}{m^2}
-\frac{5}{6} \right]\\
B_{22}(q^2,m^2)&=&-\frac{q^2}{12}
\left[\left(1-\frac{4m^2}{q^2}\right)\overline{B}
+\frac{A}{m^2}\left(1-\frac{6m^2}{q^2}\right)-\frac{1}{3}\right].
\label{b22n}
\eea

\section{Calculation for $A^8\ra\pi^+\pi^-$}
\mb{.5cm}
Now equipped with various ways to handle the integrals
appearing in our calculation, we present the relevant
details, which would be a distraction in the main body of the text.

We begin by considering Fig.~\ref{graphs}. We split the contributions
to the amplitude into $a_\mu$, $b_\mu$ and $c_\mu$ (in an obvious way).
The outgoing pions are assigned momenta $p^+$ and $p^-$ and we let $k$
be the loop momentum in amplitudes $b_\mu$ and $c_\mu$. {From}
Eq.~(\ref{chiral-current}) we know that $a_\mu=0$ to this order
in the chiral series. So we turn to
the $O(\pi^4)$ pieces of $J^8_\mu$ to determine $c_\mu$ which is given by,
\be
c_\mu=-\frac{\sqrt{3}}{4F^2}\int_k((p^+-p^-)_\mu-2k_\mu)\frac{1}{k^2-
m_{K^+}^2}-[K^+\leftrightarrow K^0],
\ee
where we have used an obvious notation for the integral over $d^Dk/(2\pi)^D$.
In dimensional regularisation, the pieces proportional to $k_\mu$
form an odd function and vanish upon integration leaving,
\be
c_\mu=-\frac{\sqrt{3}}{4F^2}(p^+-p^-)_\mu\int_k\frac{1}{k^2-
m_{K^+}^2}-[K^+\leftrightarrow K^0].
\ee
The contribution $b_\mu$ is slightly more complicated, as we have to consider
two ChPT vertices which we shall call $V_\mu$ (a four-vector)
and $S$ (a scalar). The loop integral now has two propagators,
\be
b_\mu=\int_kV_\mu\frac{1}{(k^2-m_{K^+}^2)((k+q)^2-m_{K^+}^2)}S
-[K^+\leftrightarrow K^0].
\label{bmu}
\ee
{From} Eq.~(\ref{chiral-current})
\be
V_\mu=\frac{\sqrt{3}}{2}(2k_\mu+q_\mu).
\ee
This deserves a moment's consideration. If $S$ had no $k$ dependence
then $b_\mu$ would be proportional to $q_\mu(2B_1+B)$ which vanishes
as $B_1=-B/2$ \cite{hollik}. Therefore the only parts of Eq.~(\ref{bmu}) that
will survive are those for which $S$ contains $k$. We find that
these terms are
\be
S=-\frac{1}{6F^2}(-3p^+\!\cdot\! k+3p^-\!\cdot\! k+q\!\cdot\! k+k^2).
\ee
We can now write,
\be
b_\mu=\frac{\sqrt{3}}{12F^2}
\int_k\frac{(2k_\mu+q_\mu)(3(p^+-p^-)\!\cdot\! k-(q\!\cdot\! k+k^2))
}{(k^2-m_{K^+}^2)((k+q)^2-m_{K^+}^2)}-[K^+\leftrightarrow K^0].
\label{pig}
\ee
Before attempting to evaluate this, it helps to first consider
that, because $B_1=-B/2$, we can add terms independent of $k$ to the
numerator of Eq.~(\ref{pig}), hence
\bea\non
\int_k\frac{(2k_\mu+q_\mu)(q\!\cdot\! k+k^2)}{(k^2-m^2)((k+q)^2-m^2)}
&=&\h\int_k\frac{(2k_\mu+q_\mu)(k^2-m^2+(q+k)^2-m^2)}{(k^2-m^2)((k+q)^2-m^2)}\\
&=&\h\int_k\left(\frac{2k_\mu+q_\mu}{(q+k)^2-m^2}+
\frac{2k_\mu+q_\mu}{k^2-m^2}\right)\non\\
&=&\h\int_k\left(\frac{k_\mu-q_\mu}{k^2-m^2}+
\frac{q_\mu}{k^2-m^2}\right)\non \\
&=&0.\non
\eea
So the only surviving piece of Eq.~(\ref{pig}) is (recalling
the factor $\sigma$ defined in Eq.~(\ref{sigma}))
\bea\non
b_\mu&=&\frac{\sqrt{3}}{12F^2}\int_k\frac{3k^\nu(p^+-p^-)_\nu(2k_\mu+q_\mu)
}{(k^2-m_{K^+}^2)((k+q)^2-m_{K^+}^2)}-[K^+\leftrightarrow K^0]\\
&=&\frac{\sqrt{3}}{4F^2}\sigma(p^+-p^-)^\nu(2B_{\mu\nu}(K^+)
+q_\mu B_\nu(K^+))-[K^+\leftrightarrow K^0]\non\\
&=&\frac{\sqrt{3}}{2F^2}
\sigma(p^+-p^-)^\nu B_{22}(K^+)-[K^+\leftrightarrow K^0],
\eea
as $q\!\cdot\!(p^+-p^-)=0$.

We now have the expression for the amplitude,
\bea
\M_\mu&=&b_\mu+c_\mu \non\\
&=&\frac{\sqrt{3}}{4F^2}\sigma(p^+-p^-)_\mu
(2B_{22}(K^+)-A(K^+)-[K^+\leftrightarrow
K^0]).
\label{anamp}
\eea
We now turn to Eqs.~(\ref{aeq}) and
(\ref{b22n}) to find
expressions for $A(m^2)$ and $B_{22}(q^2,m^2)$ respectively.
Substituting, we find
\be
2B_{22}(K^+)-A(K^+)-[K^+\leftrightarrow
K^0]=\frac{q^2}{6}\ln\frac{m^2_{K^+}}{m^2_{K^0}}
-\frac{q^4}{60}\left(\frac{1}{m^2_{K^+}}-\frac{1}{m^2_{K^0}}\right).
\label{simpson}
\ee
Eqs.~(\ref{anamp}) and (\ref{simpson}) can then be combined to give
us an expression for the form-factor $F_\pi^8(q^2)$, Eq.~(\ref{firstone}).

\end{document}